\documentclass[aip,twocolumn,amsmath,amssymb,reprint]{revtex4-1}

\usepackage{dcolumn}
\usepackage{bm}
\usepackage[utf8]{inputenc}
\usepackage[T1]{fontenc}
\usepackage{mathptmx}

\usepackage{amsmath} 
\usepackage{amsfonts}
\usepackage{commath}
\usepackage{url}
\usepackage{graphicx}
\usepackage{setspace}
\usepackage{soul}
\usepackage{color}

\setkeys{Gin}{width=0.9\linewidth,keepaspectratio}

\DeclareMathAlphabet{\mathpzc}{OT1}{pzc}{m}{it}

\newcommand{\pfrac}[2]{\frac{\partial #1}{\partial #2}}
\newcommand{\pfraca}[1]{\frac{\partial}{\partial #1}}

\newcommand{\mvec}[1]{\bm{#1}}

\newcommand{\script}[1]{\mathpzc{#1}}

\begin{document}

\title{A boundary value ``reservoir problem'' and boundary conditions for multi-moment multifluid simulations of sheaths}

\author{P. Cagas}%
\email{pcagas@vt.edu}
\affiliation{Virginia Tech, Blacksburg, VA, USA}%

\author{Ammar H. Hakim}
\affiliation{
Princeton Plasma Physics Laboratory, Princeton NJ 08543-0451, USA
}%

\author{B. Srinivasan}
\email{srinbhu@vt.edu}
\affiliation{Virginia Tech, Blacksburg, VA, USA}%

\date{\today}

\begin{abstract}
Multifluid simulations of plasma sheaths are increasingly used to
model a wide variety of problems in plasma physics ranging from global
magnetospheric flows around celestial bodies to plasma-wall
interactions in thrusters and fusion devices.  For multifluid
problems, accurate boundary conditions to model an absorbing wall that
resolves a classical sheath remains an open research area. This work
justifies the use of vacuum boundary conditions for absorbing walls to
show comparable accuracy between a multifluid sheath and lower moments
of a continuum-kinetic sheath.
\end{abstract}

\maketitle


Multifluid and continuum-kinetic simulations of plasma sheaths often
use absorbing walls as boundary conditions in the absence of material
emission\cite{cagas2017continuum,Cagas2020plasma} and other
plasma-wall effects. In non-neutral two-fluid modeling of sheath
dynamics, in which ions and electrons are treated as independent
interpenetrating fluids, boundary conditions need to be specified to
appropriately treat subsonic and supersonic
quantities.\cite{wilcoxson1996simulation, alvarez2020plasma} Electrons
travel across the sheath edge at bulk speeds lower than their thermal
velocity at the boundary whereas ions travel at speeds larger than the
Bohm speed.\cite{langmuir1929general, bohm1949characteristics,
  riemann1991bohm} Commonly used models for multifluid sheaths specify
a flux at the wall for the subsonic electrons based on the classical
Bohm velocity\cite{alvarez2020plasma} but this description does not
resolve the sheath profile accurately as the Bohm velocity is achieved
at the sheath entrance and not at the wall. Also, the exact form of
the flux to be specified remains somewhat ad-hoc. Hence, appropriate
sheath boundary conditions for multifluid descriptions remain an open
research area.\cite{kuhn2006link, keidar2009sheath, loizu2012boundary}

This work proposes using Riemann solvers at the plasma-material wall,
with the wall treated as a vacuum. A Riemann solver automatically
provides the needed flux for all fluid quantities at the wall,
obviating the need to explicitly specify the fluxes. Multifluid
simulations\cite{Hakim:2006iw, loverich2005discontinuous,
  shumlak2011advanced, srinivasan2011numerical,cagas2017continuum} of
classical sheaths using these Riemann solver boundary conditions are
compared to fully kinetic simulations\cite{juno2018discontinuous,
  cagas2017continuum} to show that the agreement of the lower moments
is excellent. Of course, kinetic physics (or higher-moment models that
include pressure-tensor and higher moments) is essential to accurately
model the sheath to account for anisotropic temperature, range of
collisionalities between a presheath and a sheath, critical role of
heat flux and kinetic effects in the presence of magnetic
fields.\cite{guo2012parallel, tang2016critical} Yet, multi-moment
multifluid models remain a computationally tractable reduced model.


Firstly, a generic boundary value problem (BVP) is introduced, denoted
as the ``reservoir problem'', for a system of hyperbolic
equations. Also presented is the solution to this problem in terms of
another well-known problem for hyperbolic equations, that is, the
Riemann problem. Unlike the reservoir problem, the Riemann problem is
an initial value problem (IVP) but its solution can be used to
construct a solution to the reservoir problem. The connection to
sheath boundary conditions is that the boundary where plasma touches
the material surface is a special case of the reservoir problem in
which the ``reservoir'' is a vacuum. However, setting other values for
the wall reservoir can allow modeling of other physical processes at
the wall, for example, evaporation and condensation in neutral fluid
simulations.\cite{Goldston:2017gn} In this work, "reservoir" is
synonymous to setting a ghost cell boundary condition and allowing a
self-consistent solution to the Riemann problem at the boundary.

Consider a system of 1D hyperbolic conservation laws written as
\begin{align}
  \pfrac{\mvec{Q}}{t} + \pfrac{\mvec{F}}{x} = 0,
\end{align}
where $\mvec{Q}(x,t)$ are the conserved quantities and
$\mvec{F}(\mvec{Q})$ are the fluxes. The reservoir problem can be
stated as finding the steady-state solution $\mvec{Q}^*$ to this
problem with the boundary conditions
\begin{align}
  \mvec{Q}(x<0,t) &= \mvec{Q}_L \label{eq:leftres}\\
  \mvec{Q}(x>1,t) &= \mvec{Q}_R  \label{eq:rightres}
\end{align}
in the domain $x\in[0,1]$. Note that this system is assumed to be
\emph{hyperbolic}, i.e. the flux Jacobian $\mvec{A} \equiv \partial
\mvec{F}/\partial \mvec{Q}$ is diagonalizable and has real
eigenvalues.

Consider the simplest case of linear advection,
\begin{align}
  \pfrac{q}{t} + \lambda\pfrac{q}{x} = 0.
\end{align}
For this simple case, the solution to the reservoir problem is
trivial: if $\lambda>0$ then $q^* = q_L$ and if $\lambda<0$ then $q^*
= q_R$. For linear hyperbolic \emph{systems}, for example, Maxwell
equations
\begin{align}
  \frac{\partial }{\partial t}
  \left[
    \begin{matrix}
      E_y \\
      B_z
    \end{matrix}
  \right]
  +
  \frac{\partial }{\partial x}
  \left[
    \begin{matrix}
      B_z \\
      E_y
    \end{matrix}
  \right]
  =
  0 
\end{align}
(speed of light is set to unity) one can use a diagonalization
procedure: add and subtract the two equations to get the uncoupled
system
\begin{align}
  \pfraca{t}(E_y+B_z) + \pfraca{x}(E_y+B_z) &= 0 \\
  \pfraca{t}(E_y-B_z) - \pfraca{x}(E_y-B_z) &= 0.
\end{align}
Hence, one must have $E^*_y+B_z^* = E_{L,y}+B_{L,z}$ and $E^*_y-B_z^*
= E_{R,y}-B_{R,z}$ and the solution to the reservoir problem is
\begin{align}
  E_y^* &= \frac{1}{2}\left(E_{R,y}+E_{L,y}\right) - \frac{1}{2}\left(B_{R,z}-B_{L,z}\right) \\
  B_z^* &= \frac{1}{2}\left(B_{R,z}+B_{L,z}\right) - \frac{1}{2}\left(E_{R,y}-E_{L,y}\right).
\end{align}
Note that in general there will be (potentially large) finite jumps at
the walls.

For nonlinear hyperbolic equations the situation is more
complicated. The jump between the reservoir values and the unknown
solution $\mvec{Q}^*$ will lead to a flux and for steady state the
flux must be the same at the left and right walls. The solution to the
Riemann problem can determine this flux. Consider an interface with
left and right values $\mvec{Q}^-$ and $\mvec{Q}^+$ at $t=0$. Then,
the \emph{Riemann problem} is to determine the interface value
$\tilde{\mvec{Q}}$ for $t>0$. Once this is known, then the \emph{flux}
at the interface can be computed as $\mvec{F(\tilde{\mvec{Q}})}$. Lets
denote this flux from the solution to the Riemann problem at the
interface as $\script{F}(\mvec{Q}^-,\mvec{Q}^+)$. In terms of this
flux one can determine the solution to the reservoir problem as that
value which gives
\begin{align}
  \script{F}(\mvec{Q}_L,\mvec{Q}^*) = \script{F}(\mvec{Q}^*,\mvec{Q}_R).
\end{align}
In general, this is a nonlinear system of equations that must be
solved numerically to determine $\mvec{Q}^*$. In fact, for complicated
systems like Euler equations or ideal MHD equations the Riemann solver
itself does not have an explicit analytical solution in terms of
elementary functions and must be solved partly numerically. An
additional iterative scheme is needed that calls the Riemann solver in
a loop till a converged $\mvec{Q}^*$ is found satisfying this
condition.

As an example of a nonlinear system, consider the Euler equations for
ideal, inviscid fluids. An exact Riemann solver in the inner loop of a
root finding algorithm is used to compute the solution to the
reservoir problem. For example, for reservoirs with prescribed mass
density, velocity, and pressure corresponding to the notation in
Eqs. \ref{eq:leftres} and \ref{eq:rightres}, $(\rho_L,u_L,p_L) = (1,
0, 1)$ and $(\rho_R,u_R,p_R) = (0.125, 0, 0.1)$ and gas-adiabatic
constant, $\gamma = 3$, the solution for the intermediate state is
$(\rho^*, u^*, p^*) = (0.65, 0.61, 0.27)$.

Note that for some nonlinear systems the solution $\mvec{Q}^*$ need
not be \emph{uniform}. As a simple example consider the Burgers'
equation that has a quadratic nonlinear flux. For specific choices of
left/right reservoirs ($Q_R = -Q_L$) a single discontinuity will form
in the domain. However, in this specific case a small perturbation to
one of the reservoirs will remove the discontinuity.

In a time-dependent sheath problem one only needs the flux, say at the
right wall, $\script{F}(\mvec{Q},\mvec{Q}_R)$, given $\mvec{Q}$, the
solution just to the left of the wall. Typically $\mvec{Q}_R =
0$. Often, when an exact Riemann solver is not available an
\emph{approximate} solver can be used to give reasonable estimates of
the flux at the wall. The exact Riemann solver may only be needed to
compute the wall fluxes and a faster and more tractable approximate
solver can be used to update the fluxes at interior cell edges. For
example, for the two-fluid model, with the two fluids being ions and
electrons coupled with Maxwell's equations,\cite{Hakim:2006iw,
  loverich2005discontinuous, shumlak2011advanced,
  srinivasan2011numerical} the exact Riemann solver involving both
fluid species and fields is not available and an approximate Riemann
solver must be used. For comparison, when applied on the previously
described reservoir problem, the approximate Riemann solver used here
gives the intermediate state of $(\rho^*, u^*, p^*) = (0.70, 0.54,
0.25)$.

In order to justify the appropriate multi-moment boundary conditions
for reservoir and sheath problems, a kinetic investigation is
performed in three stages.  For the first stage, non-vacuum reservoirs
are set as boundaries on both sides of the computational domain to
obtain the intermediate values for a single neutral species. Next, the
right reservoir is set to vacuum to ensure that the sonic outflow is
recovered accurately.  Lastly, the vacuum boundary condition is
applied to two-species kinetic and multi-moment plasma.  The plasma
sheath that self-consistently forms is compared between the kinetic
and two-fluid simulations.

Here, the reservoir problem is demonstrated using the kinetic
Vlasov-BGK (Bhatnagar-Gross-Krook) equations,
\begin{equation}\label{eq:vlasov}
    \pfrac{f}{t}+\mvec{v}\cdot\nabla_{\mvec{x}}f+\frac{q}{m}\left(\mvec{E}+\mvec{v}\times\mvec{B}\right)\cdot\nabla_{\mvec{v}}f = \nu\left(f_M-f\right),
\end{equation}
where $f$ is the particle distribution function, $q$ and $m$ are
charge and mass, respectively, $\nu$ is collision frequency, and $f_M$
is a Maxwellian particle distribution function which is constructed
from the first three moments of $f$. Note that the part with the
Lorentz force matters only for the multi-moment plasma case and not
for the single species neutral cases.  The equation is implemented in
the \textsc{Gkeyll} plasma simulation framework\cite{gkeyllWeb} using
the discontinuous Galerkin (DG)
method.\cite{juno2018discontinuous,hakim2020alias} The particle
distribution is discretized in each cell with a polynomial
approximation and a penalty flux is used to reconstruct an
intermediate state at cell interfaces.\cite{hakim2020alias}

Unlike the hyperbolic case of Euler equations discussed in the
previous section, the Vlasov-BGK equation contains the effects of
finite collisionality and kinetic effects, including heat-flux and
higher moments of the distribution function. For example, for finite
collisionality a boundary layer forms at the reservoir boundaries
(about a mean-free-path in thickness), but such a boundary-layer is
absent in the Euler case.

\begin{figure*}
    \centering
    \includegraphics[width=\linewidth]{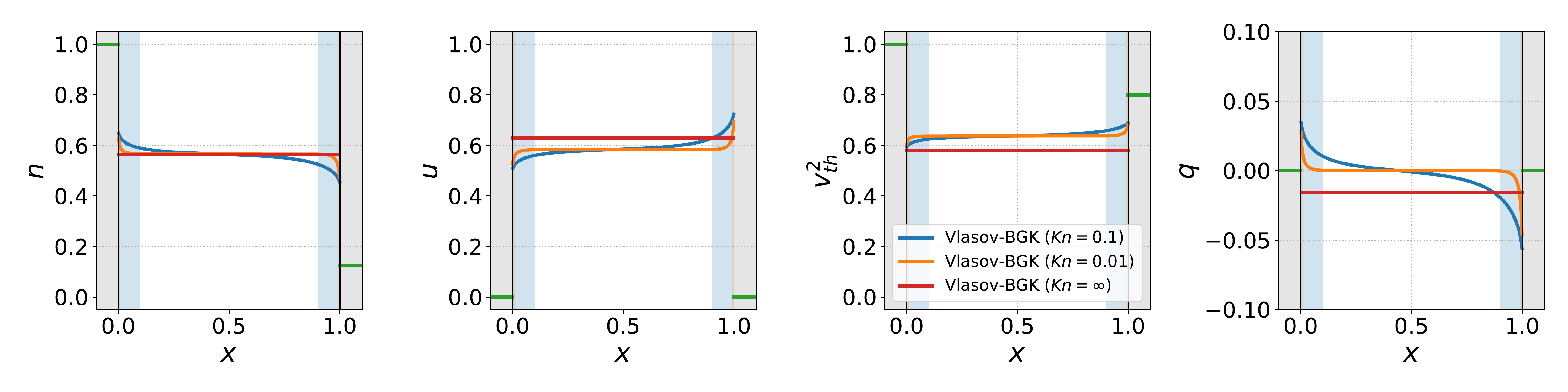}
    \caption{Spatial variation of density (left), bulk velocity
      (second from left), temperature (second from right), and heat
      flux (right) for different collisionality using the reservoir
      boundary conditions (green lines) with the Vlasov-BGK
      equations. The shaded grey regions represent the constant
      infinite reservoirs. Note the formation of boundary layers upon
      inclusion of collision, and note the importance of the third
      moment in these simulations. }
    \label{fig:reservoir}
\end{figure*}
Figure\thinspace\ref{fig:reservoir} presents the kinetic solutions for
the choice of reservoirs with Maxwellian distribution,
\begin{equation}
    f(v) = \frac{n}{\sqrt{2\pi v_{th}^2}}
    \mathrm{exp}\left(-\frac{(v-u)^2}{2v_{th}^2}\right),
\end{equation}
with $(n_L,u_L,p_L) = (1, 0, 1)$ and $(n_R,u_R,p_R) = (0.125, 0,
0.1)$, where $p=n v_{th}^2$, and collisionalities with Knudsen numbers
($\mathit{Kn}$) of $0.01$, $0.1$, and $\infty$. The density, velocity,
temperature, and the heat-flux are presented (in the fluid frame) in
Fig.\thinspace\ref{fig:reservoir}. A boundary layer forms at the walls
due to particles coming to and from the reservoir with different
velocity distributions while undergoing collisional relaxation into a
Maxwellian distribution over a finite mean-free-path. Note that this
is different from the traditional viscous boundary layer which forms
due to no-slip boundaries tangential to the wall in the Navier-Stokes
equations. As expected, for lower collisionality the heat-flux near
the wall is significant due to the non-Maxwellian shape of the
distribution.

The density of the kinetic solution can be understood by looking at
the $\mathit{Kn}=\infty$ case. For this case, the intermediate density
is the sum of two half-Maxwellians such that the half-Maxwellian with
the positive velocity is based on the conditions from the left
reservoir while the half-Maxwellian with the negative velocity is
based on conditions from the right reservoir.  The density obtained
using these two half-Maxwellians is exactly the intermediate density
value ($n=0.5625$) obtained from the kinetic solutions.

\begin{figure}
    \centering
    \includegraphics[width=\linewidth]{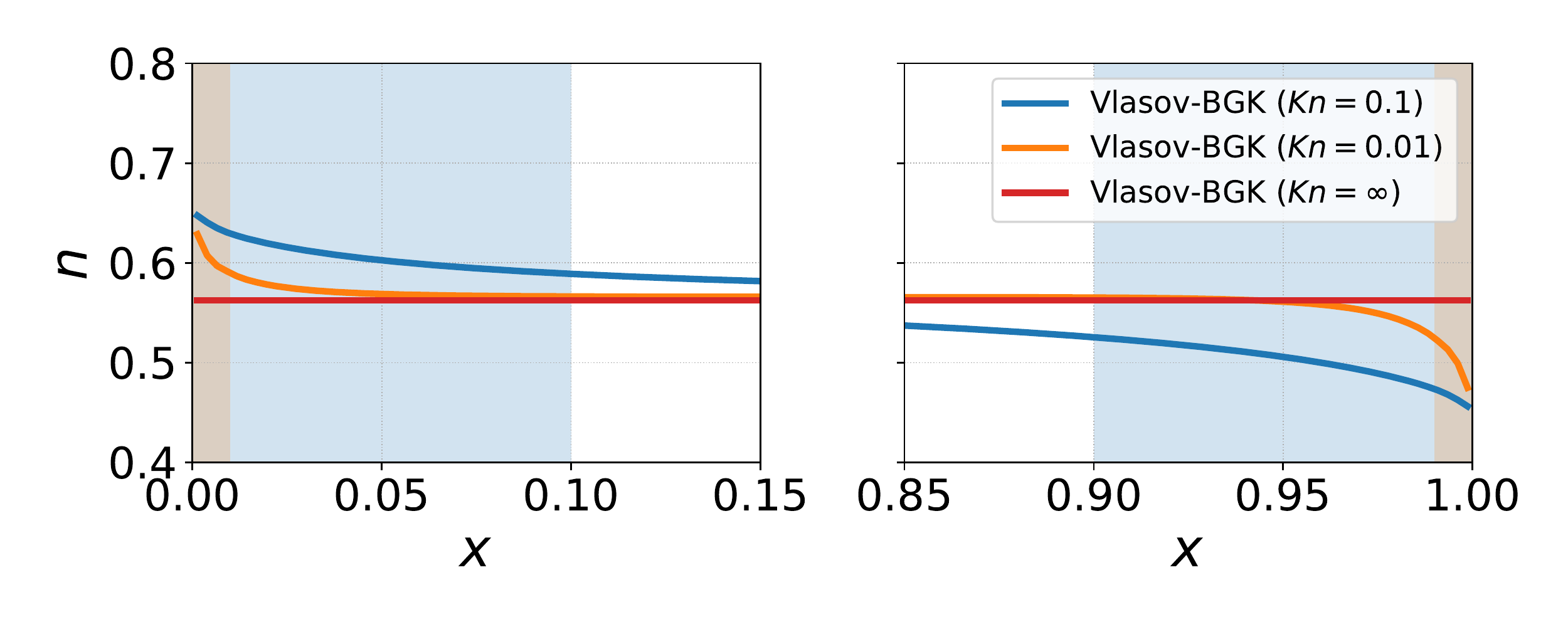}
    \caption{An expanded scale of the density in
      Fig.\thinspace\ref{fig:reservoir} for different collisionality
      using the Vlasov-BGK equations. Left plot corresponds to the
      left wall and right plot corresponds to the right wall. The
      shaded blue region marks a single mean-free-path for
      $\mathit{Kn}=0.1$ and the shaded orange region marks a single
      mean-free-path for $\mathit{Kn}=0.01$. The boundary-layer
      thickness is different at each wall as it depends not only on
      the solution in the domain, but also the reservoir levels.}
    \label{fig:reservoir_zoom}
\end{figure}

Figure\thinspace\ref{fig:reservoir_zoom} presents an expanded scale of
the kinetic solutions of Fig.\thinspace\ref{fig:reservoir} around the
left and right reservoir boundaries. The size of the boundary layer
depends on the collisionality and is denoted using the shaded
regions. The shaded regions have different colors corresponding to the
colors of the two collisional cases, and the size of the region
represents a single mean-free-path.

\begin{figure}
    \centering
    \includegraphics[width=\linewidth]{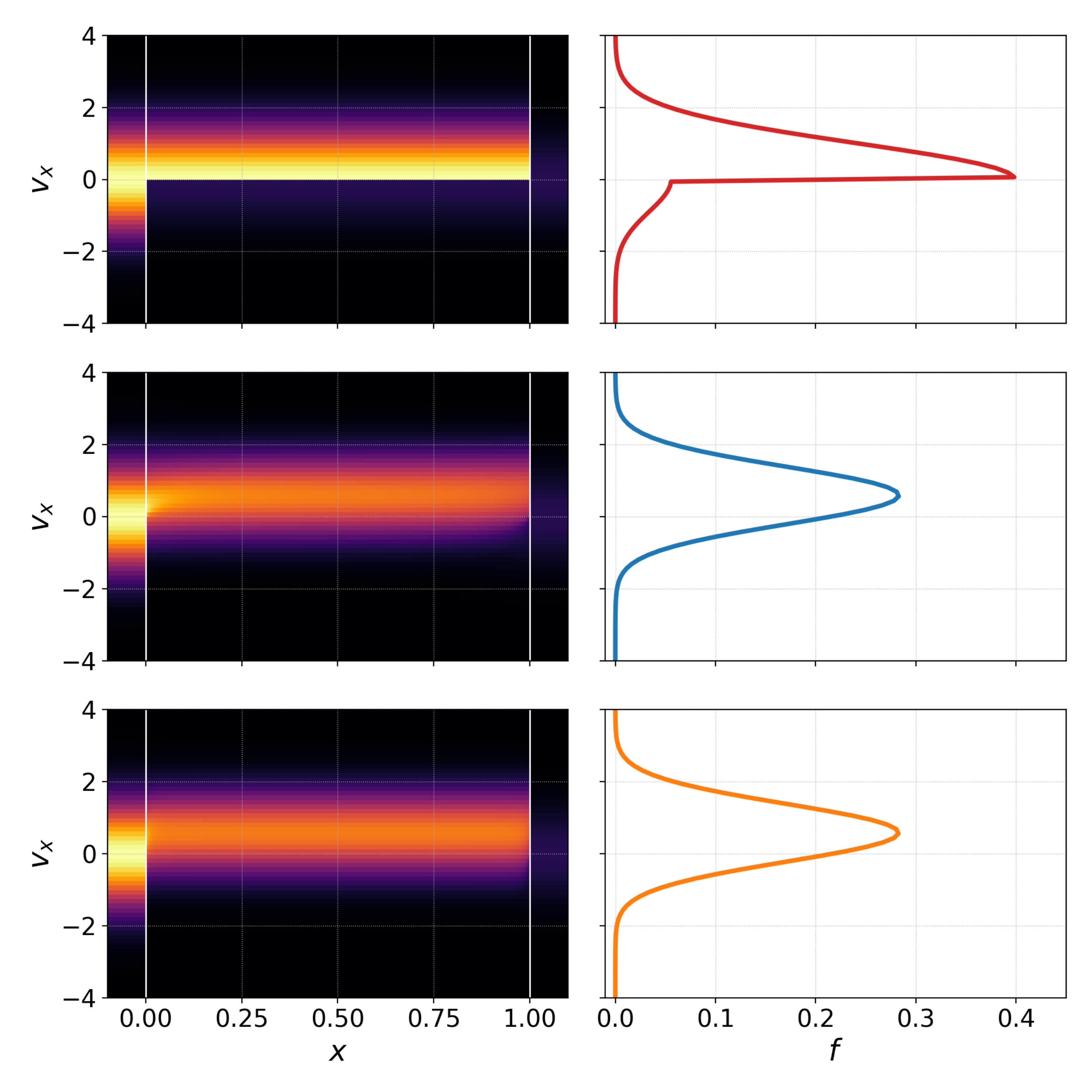}
    \caption{The distribution function (left) along with a line-out in
      the middle of the domain (right) corresponding to the three
      different collisionalities. Top plots correspond to
      $\mathit{Kn}=\infty$, middle plots correspond to
      $\mathit{Kn}=0.1$, and bottom plots correspond to
      $\mathit{Kn}=0.01$.In all the three cases, distribution
      functions near the fixed reservoirs have non-Maxwellian
      profiles, which corresponds to the non-zero heat flux seen in
      Fig.\thinspace\ref{fig:reservoir}. Note that the reservoir on
      the right side has, in these cases, much lower density and
      pressure than the left reservoir but is not zero.}
    \label{fig:reservoir_f}
\end{figure}

The kinetic distribution functions for each of the different
collisionalities are presented in Fig.\thinspace\ref{fig:reservoir_f}
along with a lineout of the distribution from the middle of the
domain. As expected, the collisionless case (top panel) represents the
half-Maxwellian distributions of each of the left and right
reservoirs. The collisionality causes thermalization of the two
Maxwellians producing the same zeroth moment (density) at the center
of the domain.  Note the expansion of the particles due to emission
and absorption near the boundaries for the collisional cases. These
boundary layers correspond to non-Maxwellian behavior and
thermalization, which also is reflected in the non-zero heat-flux at
the walls.

\begin{figure}
    \centering
    \includegraphics[width=\linewidth]{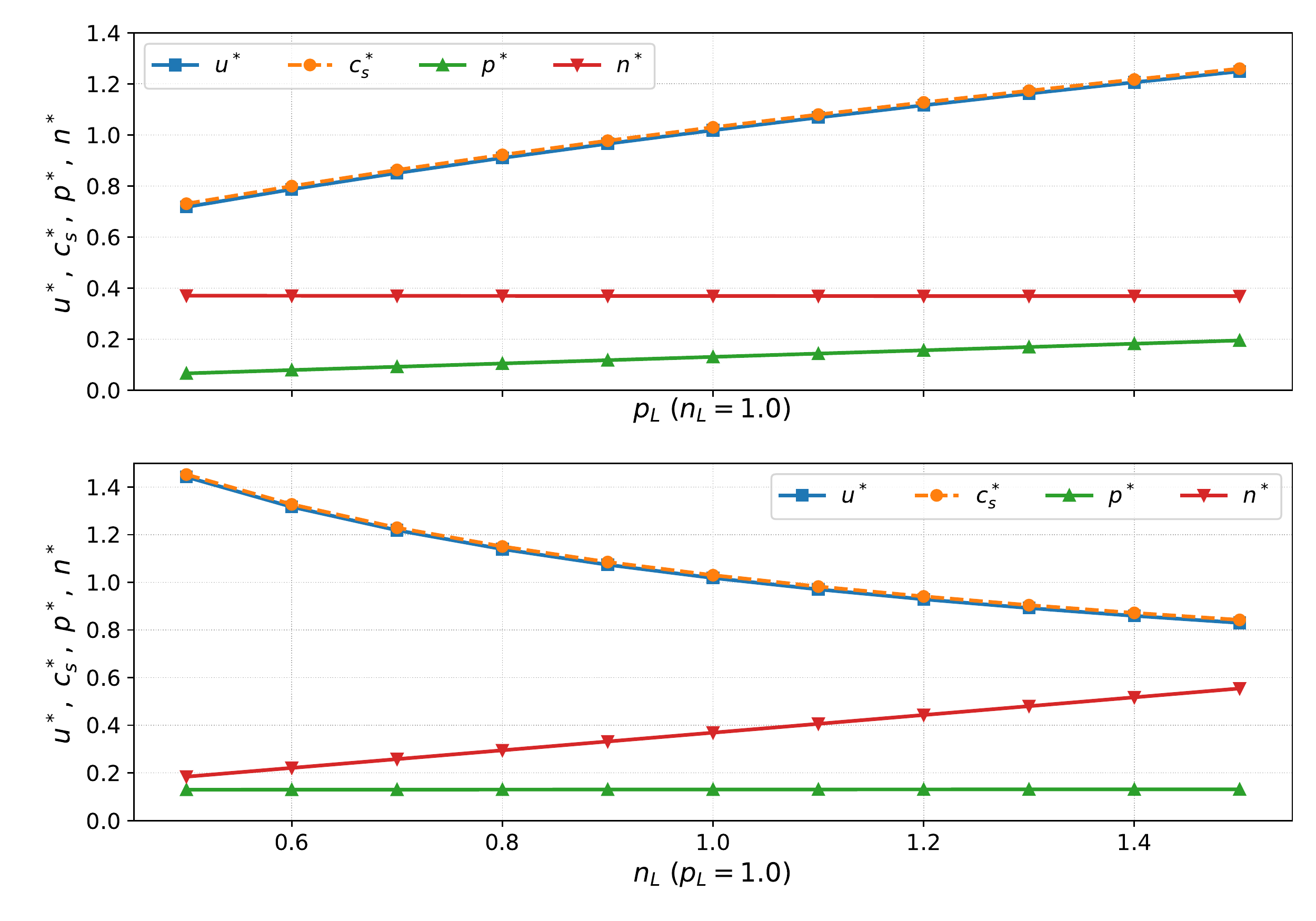}
    \caption{Kinetic simulations performed by holding the density
      fixed in the left reservoir while varying pressure (top panel)
      and holding the pressure fixed in the left reservoir while
      varying density (bottom panel). The right reservoir is a vacuum
      boundary, i.e, the distribution function, $f$, is zero. The
      plotted values are the intermediate solutions. $c_s=\sqrt{\gamma
        T/m}$ denotes the sound speed, with $\gamma=3$. Note, the
      intermediate Mach number is always $1$ as one would expect from
      outflow into a vacuum.}
    \label{fig:vac}
\end{figure}

Now consider the case where the distribution function in the right
reservoir is set to zero (equivalent to having a vacuum
reservoir). Simulations are performed by varying the left reservoir
conditions while maintaining the right reservoir as a
vacuum. Figure\thinspace\ref{fig:vac} presents kinetic results when
the left reservoir density is held fixed while varying the pressure in
the top panel and a constant pressure left reservoir with varying
density is presented in the bottom panel. Note that after reaching a
steady-state, the bulk velocity in the middle of domain exactly
matches the sound speed for both the cases.


\begin{figure}
    \centering
    \includegraphics[width=\linewidth]{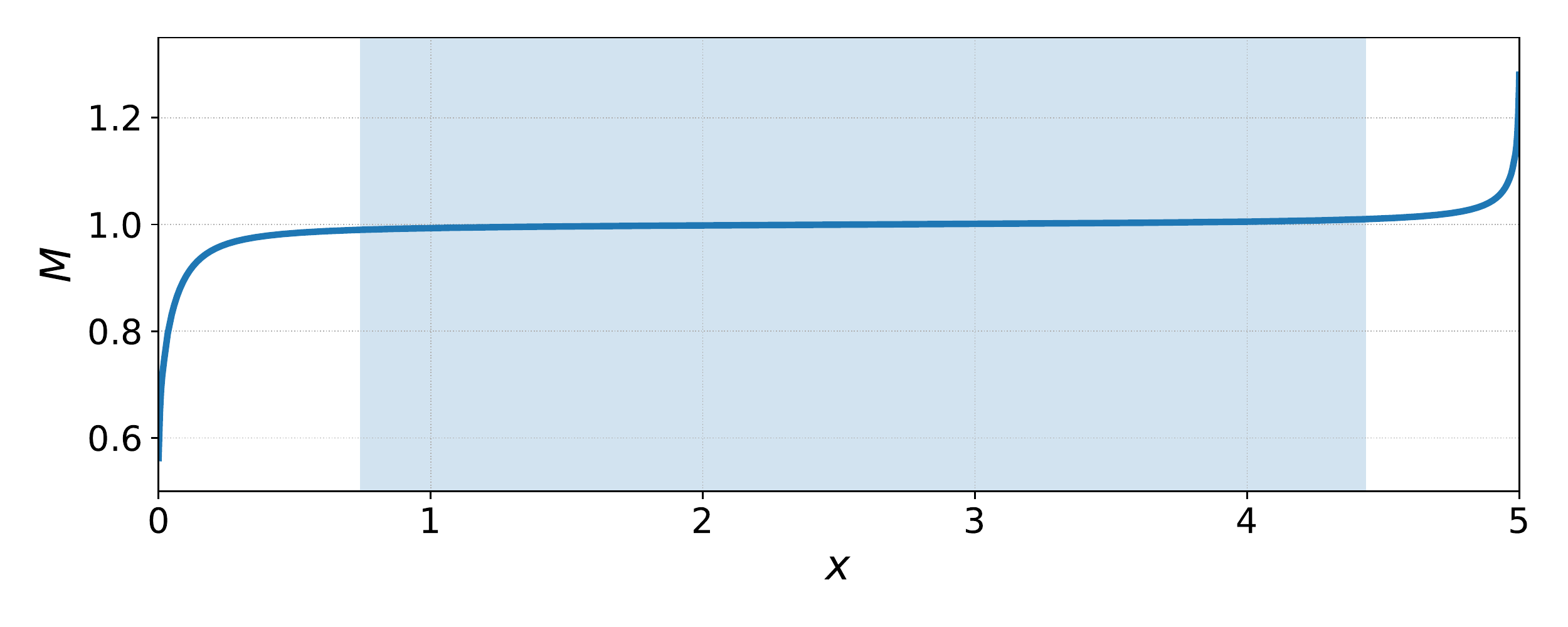}
    \caption{Mach number for $\mathit{Kn}=0.01$ for the reservoir
      conditions described in Fig.\thinspace\ref{fig:vac}. The
      particles accelerate to a sonic bulk velocity in the interior
      and then to supersonic velocity as they expand into the
      vacuum. }
    \label{fig:kin_M_5}
\end{figure}

For the collisional case of $\mathit{Kn}=0.01$,
Fig.\thinspace\ref{fig:kin_M_5} shows that the intermediate Mach
number outside of the boundary layer is exactly $1$.  The fluid
accelerates from zero to a sonic Mach across the left boundary layer,
remains sonic across most of the intermediate region, and accelerates
to a supersonic Mach number across the right boundary layer. This
occurs as the density drops across the boundary layer into the vacuum,
hence the velocity must increase to maintain constant flux across the
domain. This produces a supersonic outflow at the
  right boundary illustrating that such a vacuum boundary condition
  effectively captures an absorbing wall at the right edge of the
  computational domain.


Finally, this absorbing wall boundary condition is used to simulate a
plasma sheath.  Each species of ions and electrons are evolved using
Eq.\thinspace\ref{eq:vlasov} and coupled with Maxwell's
equations.\cite{cagas2017continuum} A classical sheath is initialized
using approximate sheath conditions presented in
Ref.[\onlinecite{cagas2017continuum},\onlinecite{robertson2013sheaths},\onlinecite{cagas2018}]. The
electron and ion populations are initialized with $T_e/T_i = 1.0$.  A
two-fluid model is also used with the two fluids being ions and
electrons coupled with Maxwell's equations.\cite{Hakim:2006iw,
  loverich2005discontinuous, shumlak2011advanced,
  srinivasan2011numerical} Analogous to kinetic simulations where the
particle distribution function coming from the wall is set to zero, a
fluid vacuum boundary condition of ($n=10^{-13}$, $p=10^{-11}$) can be
specified for each species. Maintaining a vacuum density and pressure
at the wall produces a self-consistent solution to the Riemann problem
given the upstream conditions within the plasma. The solution is then
allowed to evolve to a kinetic and two-fluid classical sheath,
respectively.

\begin{figure}
    \centering
    \includegraphics[width=\linewidth]{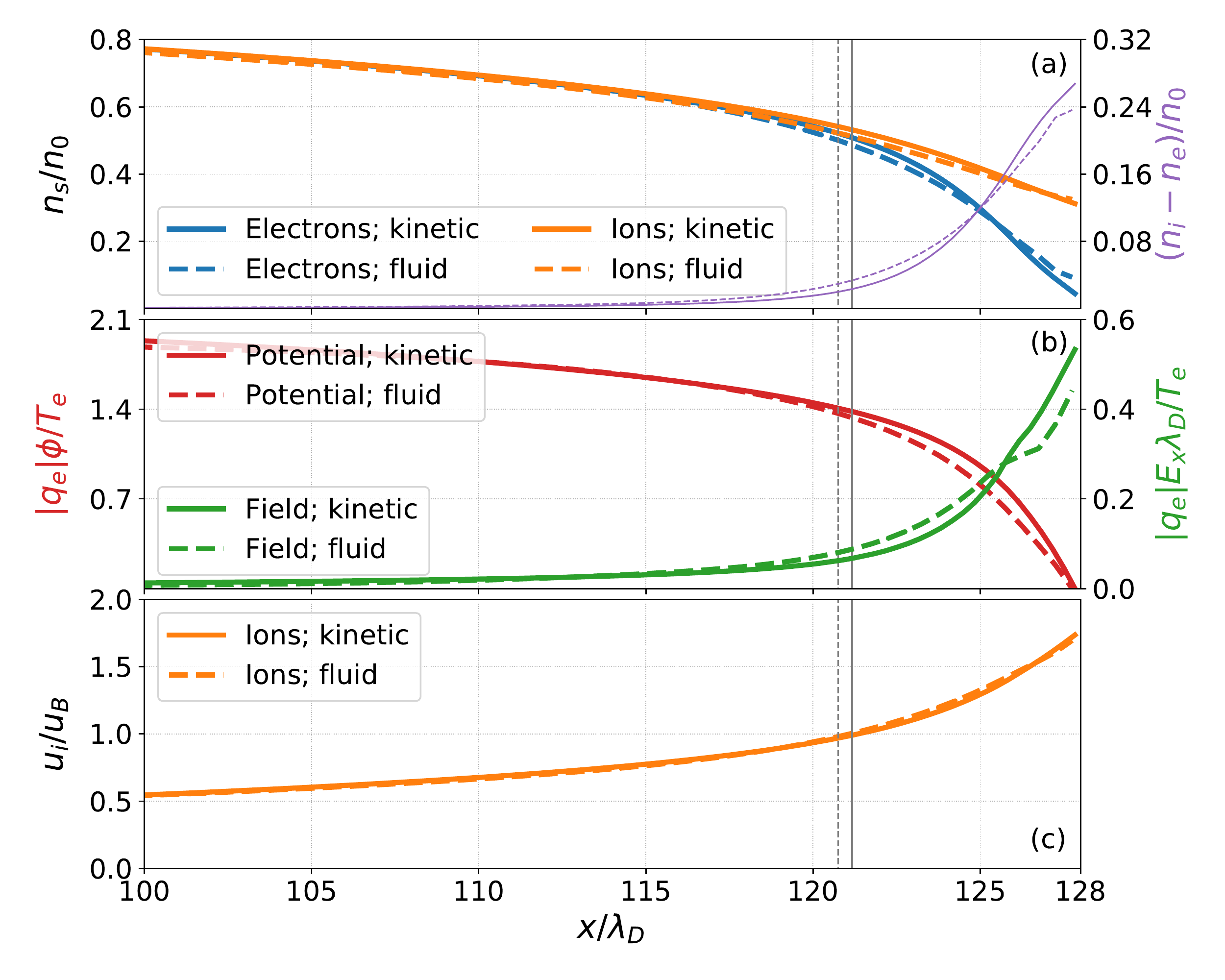}
    \caption{Comparison of electron and ion number densities (a),
      electric field (b), and ion bulk velocity (c) between kinetic
      (solid lines) and fluid (dashed lines) simulations in the region
      near an ideally absorbing wall. Vertical gray lines mark the
      crossing of the Bohm velocity. Violet lines in the panel (a)
      mark the difference between electron and ion
      densities. Initially the temperatures of the electrons and ions
      are set to $T_e/T_i = 1$. Solution is shown at $t\omega_{pe} =
      200$ and it is not averaged. Figure from Ph.D. dissertation
      [\onlinecite{cagas2018}]; used with permission.}
    \label{fig:comp_profiles}
\end{figure}

Comparisons of density, electric field, and bulk velocity profiles are
presented in Fig.\thinspace\ref{fig:comp_profiles} at a time of
$t\omega_{pe}= 200$ comparing the continuum-kinetic and two-fluid
results. The density (top), electric field (middle), and velocity
(bottom) profiles agree remarkably well between the two models. Note
that these are snapshots in time and data have not been averaged. As
this simulation was run without collisions, electron momentum
fluctuates due to Langmuir waves.\cite{Cagas:2018tz} These
fluctuations can be alleviated by averaging over the plasma frequency
to provide electron and ion fluxes that are equal to each other at the
wall with value of $0.49\,n_0 u_B$. Relying on the Riemann solver at
the absorbing boundary to self-consistently produce a plasma sheath,
the fluid simulation with the vacuum boundary condition closely
reproduces the kinetic solution.

Panel (a) also shows the deviation from quasineutrality by tracking
the absolute difference between the electron and ion number densities
normalized to the initial uniform density, as denoted by the violet
line. Note that as the electric field increases into the presheath
towards the sheath, so does the difference in the number densities,
which is proportional to the charge density.  Due to the smoothness of
the density profiles, there is no sharp transition between the
quasinetral presheath and charged sheath. An important point in the
sheath theory is the location where the ion bulk speed crosses the
classical Bohm velocity, $u_B=\sqrt{(T_e+\gamma T_i)/m_i}$. This is
marked with vertical gray lines (solid for kinetic and dashed for
fluid results).  At this point the deviation from quasineutrality is
2.3\% for the kinetic code and 2.9\% for the fluid code.

This work has shown that the Riemann problem, either approximate or
exact, provides a self-consistent boundary condition where the flux is
completely determined by prescribed vacuum regions in the boundary.
This eliminates the need for \textit{ad hoc} implementation of fluxes
that have been noted in previously used multifluid simulations of
sheaths.

To allow readers to reproduce our results and also use \textsc{Gkeyll}
for their applications, the code and input files used here are
available online. Full installation instructions are provided on the
\textsc{Gkeyll} website [\onlinecite{gkeyllWeb}].
The input files used here are under version control and can be
obtained from the repository at
\url{https://github.com/ammarhakim/gkyl-paper-inp}.

\begin{acknowledgments}
  The work presented here was supported by the U.S. Department of
  Energy Office of Science under grant number DE-SC0018276 and the
  U.S. Department of Energy ARPA-E BETHE program under grant number
  DE-AR0001263. The work of AH is also partially supported via DOE
  contract DE-AC02-09CH11466 for the Princeton Plasma Physics
  Laboratory.
\end{acknowledgments}

\appendix

\bibliography{gkyl}

\end{document}